\documentstyle[12pt,epsfig]{article}
 1
 1
 1

\begin{document}
\bibliographystyle{unsrt}
%
\def\lta{\;\raisebox{-.5ex}{\rlap{$\sim$}} \raisebox{.5ex}{$<$}\;}
\def\gta{\;\raisebox{-.5ex}{\rlap{$\sim$}} \raisebox{.5ex}{$>$}\;}
\def\grle{\;\raisebox{-.5ex}{\rlap{$<$}}    \raisebox{.5ex}{$>$}\;}
\def\legr{\;\raisebox{-.5ex}{\rlap{$>$}}    \raisebox{.5ex}{$<$}\;}

%
\def\r#1{\ignorespaces $^{\rm #1}$} 
\def\l#1{\ignorespaces $_{\rm #1}$} 

\newcommand{\ra}{\rightarrow}
\newcommand{\permille}{$^0 \!\!\!\: / \! _{00}$}
\newcommand{\dd}{{\rm d}}
\newcommand{\oal}{{\cal O}(\alpha)}%
\newcommand{\su}{$ SU(2) \times U(1)\,$}
 
\newcommand{\np}{Nucl.\,Phys.\,}
\newcommand{\pl}{Phys.\,Lett.\,}
\newcommand{\pr}{Phys.\,Rev.\,}
\newcommand{\prl}{Phys.\,Rev.\,Lett.\,}
\newcommand{\prep}{Phys.\,Rep.\,}
\newcommand{\zp}{Z.\,Phys.\,}
\newcommand{\sovjnp}{{\em Sov.\ J.\ Nucl.\ Phys.\ }}
\newcommand{\nuclinst}{{\em Nucl.\ Instrum.\ Meth.\ }}
\newcommand{\annp}{{\em Ann.\ Phys.\ }}
\newcommand{\intjmp}{{\em Int.\ J.\ of Mod.\  Phys.\ }}
 
\newcommand{\eps}{\epsilon}
\newcommand{\mw}{M_{W}}
\newcommand{\mww}{M_{W}^{2}}
\newcommand{\mbb}{m_{b \bar b}}
\newcommand{\mcc}{m_{c \bar c}}
\newcommand{\mh}{m_{H}}
\newcommand{\mhh}{m_{H}^2}
\newcommand{\mz}{M_{Z}}
\newcommand{\mzz}{M_{Z}^{2}}

\newcommand{\lra}{\leftrightarrow}
\newcommand{\tr}{{\rm Tr}}
 
\newcommand{\ie}{{\em i.e.}}
\newcommand{\cm}{{{\cal M}}}
\newcommand{\cl}{{{\cal L}}}
\def\Ww{{\mbox{\boldmath $W$}}}  
\def\B{{\mbox{\boldmath $B$}}}         
\def\nn{\noindent}

\newcommand{\sinsq}{\sin^2\theta}
\newcommand{\cossq}{\cos^2\theta}
\newcommand{\be}{\begin{equation}}
\newcommand{\beq}{\begin{equation}}
\newcommand{\eeq}{\end{equation}}
\newcommand{\ee}{\end{equation}}
\newcommand{\ba}{\begin{eqnarray}}
\newcommand{\ea}{\end{eqnarray}}
\newcommand{\beqn}{\begin{eqnarray}}
\newcommand{\eeqn}{\end{eqnarray}}
\newcommand{\bea}{\begin{eqnarray}}
\newcommand{\ena}{\end{eqnarray}} 
\newcommand{\eea}{\end{eqnarray}}

\newcommand{\nl}{\nonumber \\}
\newcommand{\eqn}[1]{Eq.(\ref{#1})}
\newcommand{\ibidem}{{\it ibidem\/},}
\newcommand{\into}{\;\;\to\;\;}
\newcommand{\wws}[2]{\langle #1 #2\rangle^{\star}}
\newcommand{\smod}{\tilde{\sigma}}
\newcommand{\dilog}[1]{\mbox{Li}_2\left(#1\right)}
\newcommand{\umu}{^{\mu}}
\newcommand{\cjg}{^{\star}}
\newcommand{\lgn}[1]{\log\left(#1\right)}
\newcommand{\si}{\sigma}
\newcommand{\sit}{\sigma_{tot}}
\newcommand{\sqs}{\sqrt{s}}
\newcommand{\sih}{\hat{\sigma}}
\newcommand{\sith}{\hat{\sigma}_{tot}}
\newcommand{\p}[1]{{\scriptstyle{\,(#1)}}}
\newcommand{\res}[3]{$#1 \pm #2~~\,10^{-#3}$}
\newcommand{\rrs}[2]{\multicolumn{1}{l|}{$~~~.#1~~10^{#2}$}}
\newcommand{\err}[1]{\multicolumn{1}{l|}{$~~~.#1$}}
\newcommand{\ru}[1]{\raisebox{-.2ex}{#1}}
\newcommand{\epem}{$e^{+} e^{-}\;$}
\newcommand{\epemt}{e^{+} e^{-}\;}
\newcommand{\eeah}{$e^{+} e^{-} \ra H \gamma \;$}
\newcommand{\eahnw}{$e\gamma \ra H \nu_e W$}

\newcommand{\thebb}{\theta_{b-beam}}
\newcommand{\pte}{p^e_T}
\newcommand{\ptH}{p^H_T}
\newcommand{\gag}{$\gamma \gamma$ }
\newcommand{\gam}{\gamma \gamma }

\newcommand{\aatoh}{$\gamma \gamma \ra H \;$}
\newcommand{\egam}{$e \gamma \;$}
\newcommand{\eat}{e \gamma \;}
\newcommand{\eaeh}{$e \gamma \ra e H\;$}
\newcommand{\eaehb}{$e \gamma \ra e H \ra e (b \bar b)\;$}
\newcommand{\egebb}{$e \gamma (g) \ra e b \bar b\;$}
\newcommand{\egecc}{$e \gamma (g) \ra e c \bar c\;$}
\newcommand{\eaebb}{$e \gamma \ra e b \bar b\;$}
\newcommand{\eaecc}{$e \gamma \ra e c \bar c\;$}
\newcommand{\aah}{$\gamma \gamma H\;$}
\newcommand{\zah}{$Z \gamma H\;$}
\newcommand{\pe}{P_e}
\newcommand{\pg}{P_{\gamma}}
\newcommand{\delbb}{\Delta m_{b \bar b}}

\begin{titlepage}
\rightline{ROME1-1178/97}
\rightline{July 1997}
\vskip 22pt 

\noindent
\begin{center}
{\Large \bf Measuring {\boldmath \zah} vertex effects \\
at {\boldmath \egam} linear colliders. 
}
\end{center}
\bigskip

\begin{center}
{\large 
E.~Gabrielli~$^{a},$
$\;\,$V.A.~Ilyin~$^b \;\,$
and $\;\,$B.~Mele~$^c$. 
} \\
\end{center}

\bigskip\noindent
$^a$ University of Notre Dame, IN, USA \\
\noindent
$^b$ Institute of Nuclear Physics, Moscow State University, Russia \\
\noindent
$^c$ INFN, Sezione di Roma 1 and Rome University ``La Sapienza", Italy
\bigskip
\begin{center}
{\bf Abstract} \\
\end{center}
{\small 
The one-loop process \eaeh, for intermediate Higgs masses is considered.  
Exact cross sections for unpolarized and longitudinally
polarized beams are computed and found to be 
more than two orders of magnitude larger than the rates 
for the crossed process \eeah, in the  energy range $\sqs=(0.5\div 2)$ TeV.
We show that, apart from being  competitive with the \aatoh process 
for testing the one-loop \aah  vertex,
by requiring a final electron tagged at large angle 
the channel  \eaeh 
provides an excellent way of testing the \zah vertex. 
}
\vskip 22pt 
\vfil
\noindent
e-mail: \\
egabriel@wave.phys.nd.edu, $\;$
ilyin@theory.npi.msu.su, $\;$
mele@roma1.infn.it \\ 
\vskip 16pt 
\vfil
\noindent
To appear in the Report DESY 97-123E
\end{titlepage}
Possible ways to test the one-loop couplings $ggH$, \aah and \zah have been 
extensively studied in the literature.
Because of the nondecoupling properties of the Higgs boson, 
these vertices are sensitive
to the contribution of new particles circulating in the loops,
even in the limit $M_{new}\gg \mh$ \cite{higg}.
A measurement of the \aah and \zah  couplings
should be possible by the determination of the BR's for the decays
$H\ra \gamma \gamma$ \cite{haau,haad} and $H\ra \gamma Z$
\cite{hzad,haad}, respectively. This is true
only for an intermediate-mass Higgs boson (i.e., for 90GeV$\lta \mh
\lta 140$ GeV), where both BR($H\ra \gamma \gamma$) and
BR($H\ra \gamma Z$) reach their maximum values, which is ${\cal O}(10^{-3})$.

Another promising way of measuring the \aah coupling for an
intermediate-mass Higgs boson will be realized through the
Higgs production in \gag collisions \cite{aahu,aahd}. To this end,
the capability of tuning the \gag c.m. energy on the Higgs mass,
through  a good degree of the photons  monochromaticity, will be crucial
for not diluting too much  the $\gam \to H$ 
resonant cross section over the c.m. energy spectrum.

In this short note, we sum up the main results recently obtained 
on the Higgs production in \egam collisions 
through the one-loop process \eaeh \cite{noii}.
This channel  turns out to be an excellent means to test 
both the \aah and \zah one-loop couplings with relatively
high statistics, without requiring a fine tuning of the c.m. energy. 
While the $\gamma$-exchange
\aah contribution is dominant in the total
cross section,  by requiring a large transverse momentum of the final
electron (or Higgs boson), one enhances the $Z$-exchange \zah contribution,
while keeping the corresponding rate still to an observable level. 
The further contribution given by the box diagrams with $W$ and $Z$ exchange 
survives  at large angles too, but is relatively less important.
Furthermore, while the  \aah and \zah channels increase logarithmically with
the c.m. collision energy, the contribution from 
boxes starts decreasing at $\sqs\gta 400$ GeV.

In our study we assumed that the initial photon beam is to a
good degree monochromatic, and has  an integrated luminosity of 
 ${\cal O}(100)$ fb$^{-1}$. 

The cross section for the process \eaeh has previously been studied in the 
Weizs\"acker-Williams (WW) approximation \cite{wewi}, where the only channel
contributing is the (almost real) $\gamma$-exchange in the $t$-channel, 
induced by the \aah vertex \cite{ebol}. 
This method provides a rather good 
estimate of the \eaeh total cross sections, but it is unable 
to assess the importance of the \zah (and box) effects. 
This we will address particularly in our exact treatment of \eaeh.

Although, the cross sections for the process \eaeh are quite 
large also for heavy Higgs masses, we will concentrate on the intermediate 
Higgs mass case (hence,  assuming that the decay $H\to b \bar b$ is dominant).
  
The crossed process, \eeah, has been  studied in different papers 
\cite{barr,abba,djou}.
However, the \eeah channel suffers from small rates, which 
are further depleted at large energies by the $1/s$ behavior of the dominant
s-channel diagrams. 
As a consequence, if a $e\gamma$ option of the linear collider will be
realized with similar luminosity of the \epem option,
the \eaeh channel will turn out to be much more interesting than the 
process \eeah for finding possible deviations from  the standard-model 
one-loop Higgs vertices.

\begin{figure}
\begin{center}
\begin{picture}(160,100)
\put(-30,-70){\epsfxsize=20cm \leavevmode \epsfbox{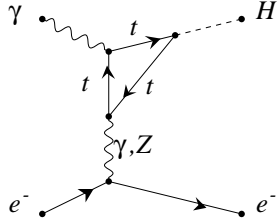} }
\end{picture} 
\end{center}
\caption{Feynman diagram with fermion triangle loop.}
\label{dgrm1}  
\end{figure}

\begin{figure}
\begin{picture}(160,100)
\put(0,-70){\epsfxsize=20cm \leavevmode \epsfbox{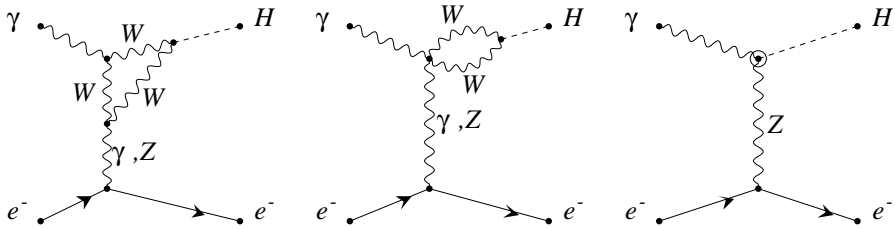} }
\end{picture} 
\caption{Feynman diagrams with $W$-triangle loop.}
\label{dgrm2}  
\end{figure}

\begin{figure}
\begin{picture}(160,100)
\put(-40,-70){\epsfxsize=20cm \leavevmode \epsfbox{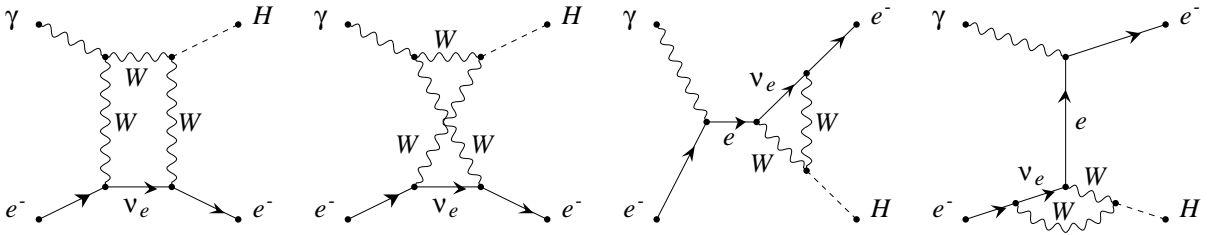} }
\end{picture} 
\caption{Subset of Feynman diagrams with $W$-box loop and related $eeH$ vertex.}
\label{dgrm3}  
\end{figure}

\begin{figure}
\begin{picture}(160,100)
\put(0,-70){\epsfxsize=20cm \leavevmode \epsfbox{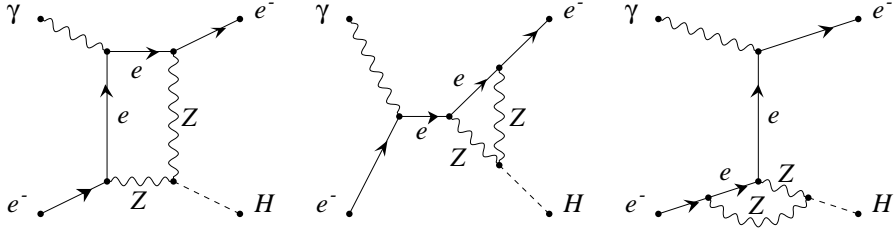} }
\end{picture} 
\caption{Subset of diagrams with $Z$-box loop and related $eeH$ vertex.}
\label{dgrm4}  
\end{figure}

In \cite{noii},
we present the complete analytical results for the helicity 
amplitudes of the \eaeh process (see also \cite{cott}). 
Although, we have calculated the relevant amplitudes for the
process \eaeh in the 't-Hooft-Feynman gauge, for brevity
we show the corresponding Feynman diagrams in the unitary gauge 
(figures~\ref{dgrm1}--\ref{dgrm4}).

\begin{figure}
\begin{picture}(160,100)
\put(-40,-70){\epsfxsize=20cm \leavevmode \epsfbox{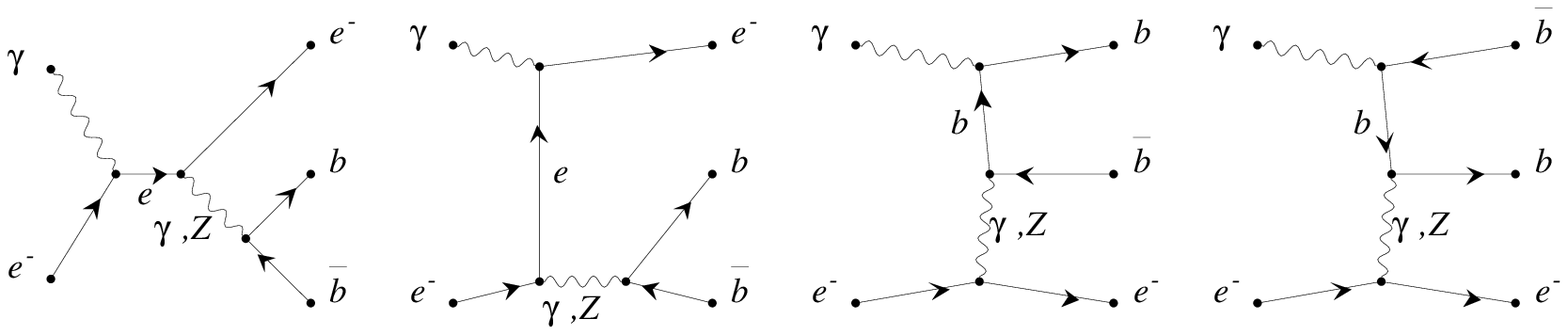} }
\end{picture} 
\caption{Diagrams for the \eaebb background.}
\label{dgrm5}  
\end{figure}
\begin{figure*}[t]
\vspace*{-4.cm}
\begin{center}
 \mbox{\epsfxsize=16cm\epsfysize=18.5cm\epsffile{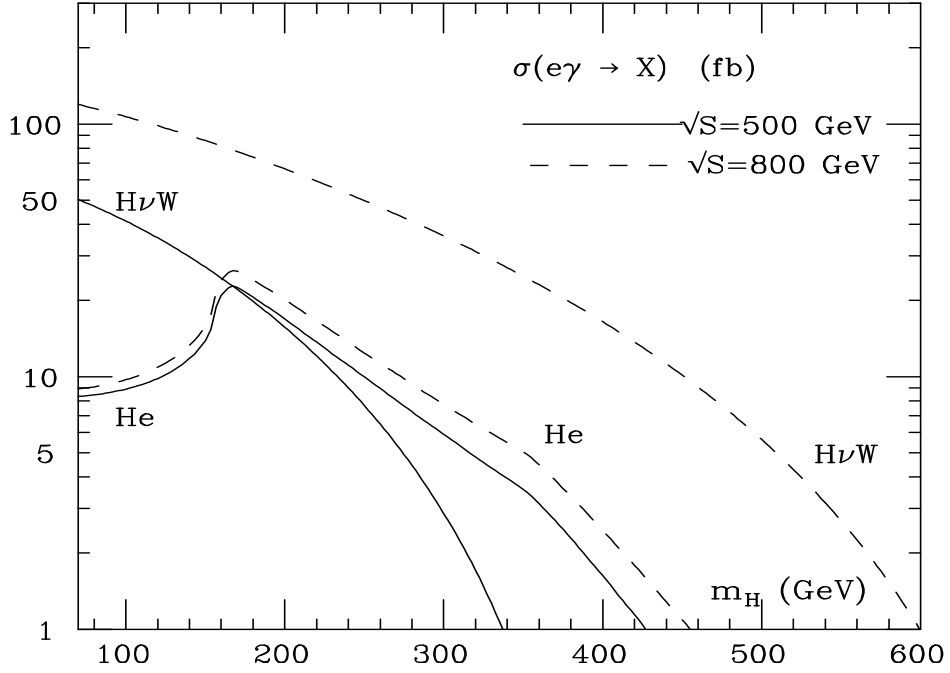}}
\vspace*{-5.5cm}
\caption{ Total cross sections for the two main H production processes.
 }
\label{fig31bis}
\end{center}
\end{figure*}

In figure \ref{fig31bis}, the total (unpolarized)
cross sections for the one-loop process \eaeh  
and the tree-level Higgs production \eahnw (computed by CompHEP \cite{comp})
are plotted versus $\mh$, for $\sqs=500$ and 800 GeV.
One can see that the process \eaeh is characterized by relatively large
rates. For instance, for $\mh$ up to about 400 GeV, one finds
$\sigma(H e) >1$ fb, which, for an integrated luminosity of about
100 fb$^{-1}$, corresponds to more than 100 events.

\begin{figure*}[t]
\vspace*{-3.5cm}
\begin{center}
 \mbox{\epsfxsize=11cm\epsfysize=14.cm\epsffile{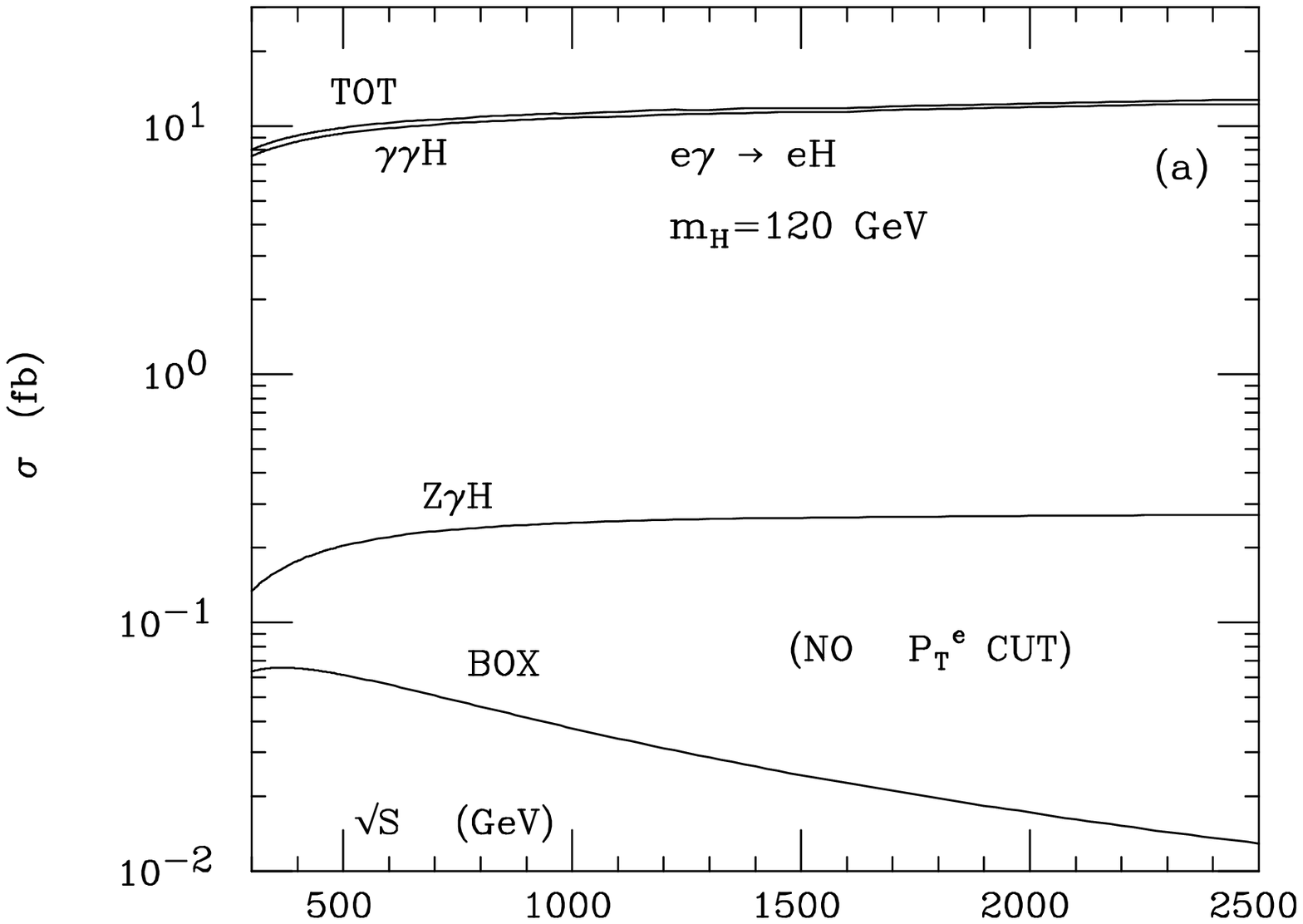}}
\end{center}
\vspace*{-8. cm}
\begin{center}
 \mbox{\epsfxsize=11cm\epsfysize=14.cm\epsffile{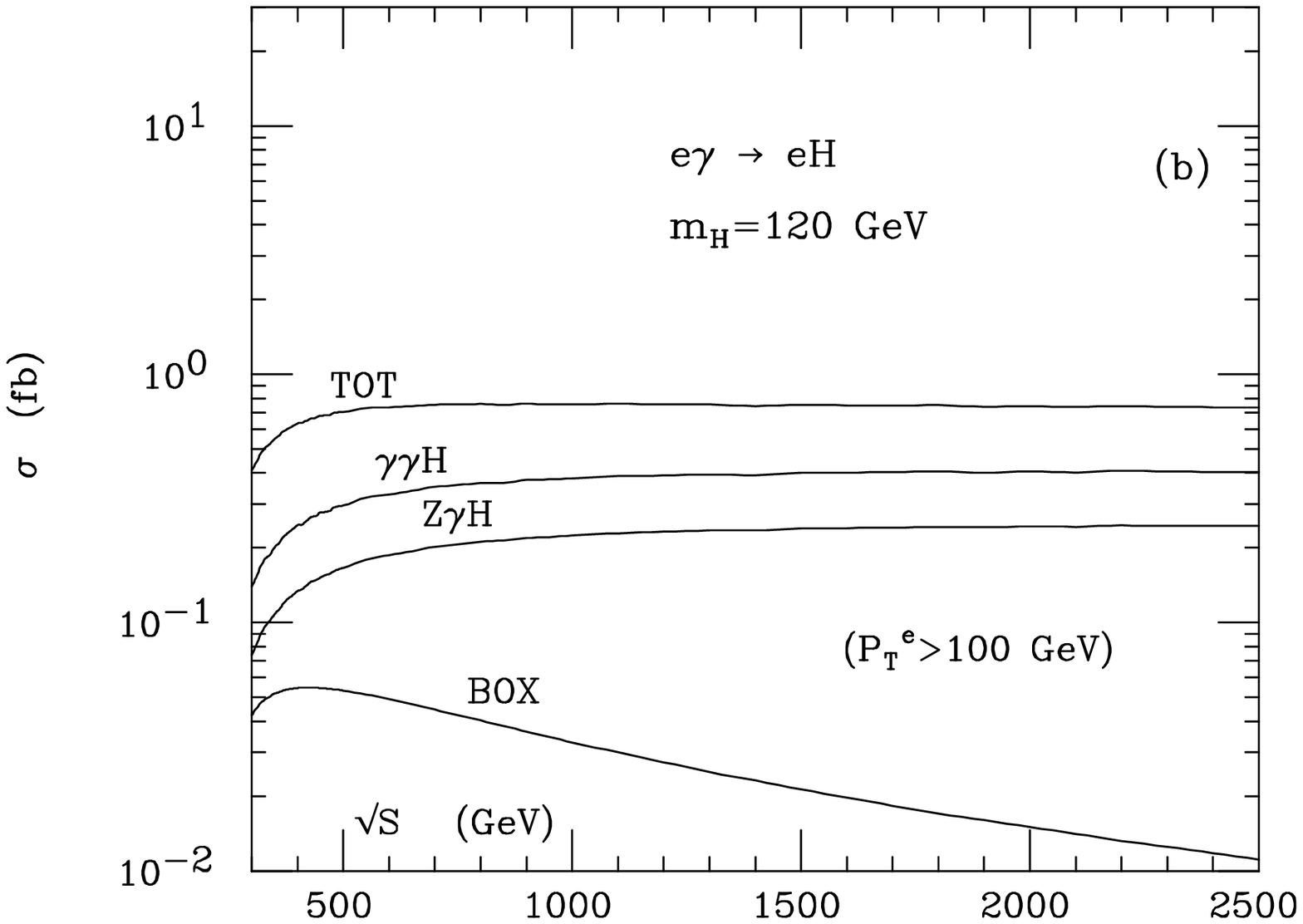}}
\vspace*{-4.cm}
\caption{ Effect of imposing a $\pte$ cut on the \eaeh cross section.}
\label{fig34}
\end{center}
\end{figure*}

A possible strategy to enhance 
the \zah vertex effects (depleted by the $Z$ propagator)
with respect to the dominant \aah contribution
in the $He$ production rate
 consists in requiring a final electron  tagged
at large angle. The corresponding cut on the transferred
squared momentum  $t$ depletes mainly the amplitudes involving a 
photon propagator in the $t$ channel. 
This can be easily seen from the two plots
in figure \ref{fig34}, where the cross sections dependence on $\sqs$,
for $\mh=120$GeV, is shown for no cut on the electron transverse
momentum $\pte$ (a), and a cut  $\pte>100$GeV  (b).
 We show separately the contributions
to the cross section  given by the squared amplitudes
corresponding to the subsets of Feynman diagrams ``\aah", ``\zah" and ``BOX"
(as defined in \cite{noii}).
Even if this separation is by no means formally rigorous
(and neglects the relative interference effects), it can help 
in getting a feeling of the relative importance of the
triangular vertices and
box contributions to the total cross section.
One can see that
the relative weight of the \zah and BOX contributions 
with respect to the total cross
section is considerably enhanced by a cut on the minimum allowed $\pte$.
For $\pte>100$GeV, \zah is about 60\% of \aah, and \zah gives
a considerable fraction of the total production rate, which still is
sufficient to guarantee investigation (about 0.7 fb).
One can also notice that the BOX contribution is of some relevance
only in the lower $\sqs$ range. 

Note that in the inclusive $He$ production, 
the bulk of the events are characterized by a forward final 
electron escaping detection. On the other hand, requiring 
a large $\pte$ corresponds, from an experimental point of view, 
to selecting a different final-state configuration, 
where the Higgs decay products have a large total transverse momentum,
balanced  by a high-energy  electron detected at large angle.

In principle, the \eaeh total cross section 
(and its main contribution from \aah)
is of the same order of magnitude of the total rates  
for Higgs production in \gag collisions \cite{aahd}.
Indeed, the expected resolution on the beam energy
smears the higher peak cross section over a width
much larger than the Higgs resonance.
As a result, the channel \eaeh has a comparable potential with
respect to the
process $\gam\to H$ in testing the \aah vertex, as far as 
the production rates are concerned.
In our work, on the other hand, we concentrated
on the problem of disentangling the \zah vertex effects, which are out of the
\gag-collision domain. 

The main irreducible background to the process
\eaehb comes from the channel \eaebb. 
The corresponding set of Feynman diagrams is given by
8 graphs and is shown in figure \ref{dgrm5}.
A crucial parameter to set the importance of the \eaebb background
is the experimental resolution on the $b\bar b$ invariant mass
$\delbb$. The background rates we present here are obtained
by integrating the $\mbb$ distribution over the range 
$\mh-\delbb<\mbb<\mh+\delbb$, assuming $\delbb=3$GeV.
We used CompHEP to generate the
\eaebb kinematical distributions and cross sections. 
The background rates turn out to be
considerably larger than the signal, especially at 
moderate values of $\pte$. The situation can be  improved 
by putting a cut on the angles between each $b$ and the initial
beams. In fact, the vector couplings that characterize
the $b$'s in the channel \eaebb give rise to a $b$ angular
distribution considerably more forward-backward peaked than
in the case of the scalar $Hb\bar b$ coupling relevant for the signal.

In table 1, the signal and background rates 
are reported, when an angular cut $\thebb>18^o$
is applied between each $b$ quark and both the beams
for two different sets of $\pte$ cuts, that enhance the \zah
contribution, and $\mh=$120 GeV, at $\sqs=$500 GeV.
The assumed $\thebb$ cut reduces the signal and background 
distributions at a comparable level, without penalizing appreciably the
signal rate at large $\pte$. Note that the cut $\thebb>18^o$
has been optimized at $\sqs=500$GeV. Lower angular cuts will be
more convenient at larger $\sqs$.
Different initial polarizations for the $e$ beam are also considered
(see below).
 
A further source of background for the process \eaehb is the charm production
through \eaecc, when the $c$ quarks are misidentified
into $b$'s. This reducible background can be cured by a good $b$-tagging
efficiency, that should control a charm production rate that can be even more
than a factor 10 larger than the corresponding \eaebb cross section,
depending on the particular kinematical configuration
\cite{ebol}.
We computed the rate for \eaecc. By assuming a 10\% probability of
misidentifying a $c$ quark into a $b$  (hence, considering only a 
fraction 1/10 of the computed \eaecc rate), 
we find that this reducible background has lower rates
than the irreducible one. This can be also  seen in table 1.
Note that the \eaecc channel is kinematically similar to
\eaebb. Hence, the particular strategies analyzed here to reduce the latter
authomatically depletes also the former. 
For unpolarized beams and $\pte>100$GeV,
the \eaecc ``effective rate" is less than 1/3 of the \eaebb rate.

A further background considered in \cite{noii}, is the resolved
\egebb production, where the photon interacts via its
gluonic content. 
Although the gluon distribution of the photon is presently poorly known, 
this possible background has also been investigated, and
should not be competitive with the previous channels.

One of the advantages of a linear collider is the possibility
to work with polarized beams. This may allow, on the one hand, to test 
the parity structure of the interactions governing a particular
process and, on the other
hand, to optimize its background suppression.
Hence, we considered the possibility of having either 
the electron or the photon beam longitudinally polarized.
In table 2, the $e/\gamma$ polarization dependence of 
the total cross section, its \aah, \zah and BOX components,
and  the  interference 
pattern of the \aah, \zah and BOX contributions
are shown for $\pte>100$GeV,  at $\sqs=500$GeV and different
values of $\mh$.
The total rates turns out to be very sensitive to the electron
polarization, in the high $\pte$ sector of the
phase-space.
For instance, assuming $\pe=-1$ ($\pe=+1$)  the total cross section
increases (decreases) by about 94\% at $\sqs=500$GeV.
For $\pe=+1$ there is a strong
destructive interference between the terms \aah and \zah.
This is essentially due to the different sign of the 
couplings $ee\gamma$ and $e_Re_RZ$, where $e_R$ stands 
for the right-handed electron component.

In table 1, one can also see that, although both 
the signal S and background B are
increased by a left-handed $e$ polarization, the ratio $S/B$ is improved at 
large $\pte$. 

The effect of a longitudinally polarized photon beam 
is similar to the $e$ polarization, but 
is quantitatively more modest, 
especially at large values of $\sqs$. 
One exception is given by  the BOX contribution
that is still considerably altered by $\pg\neq 0$ at 
any $\sqs$ \cite{noii}.

\begin{figure}[h]
\centerline{
\epsfig{figure=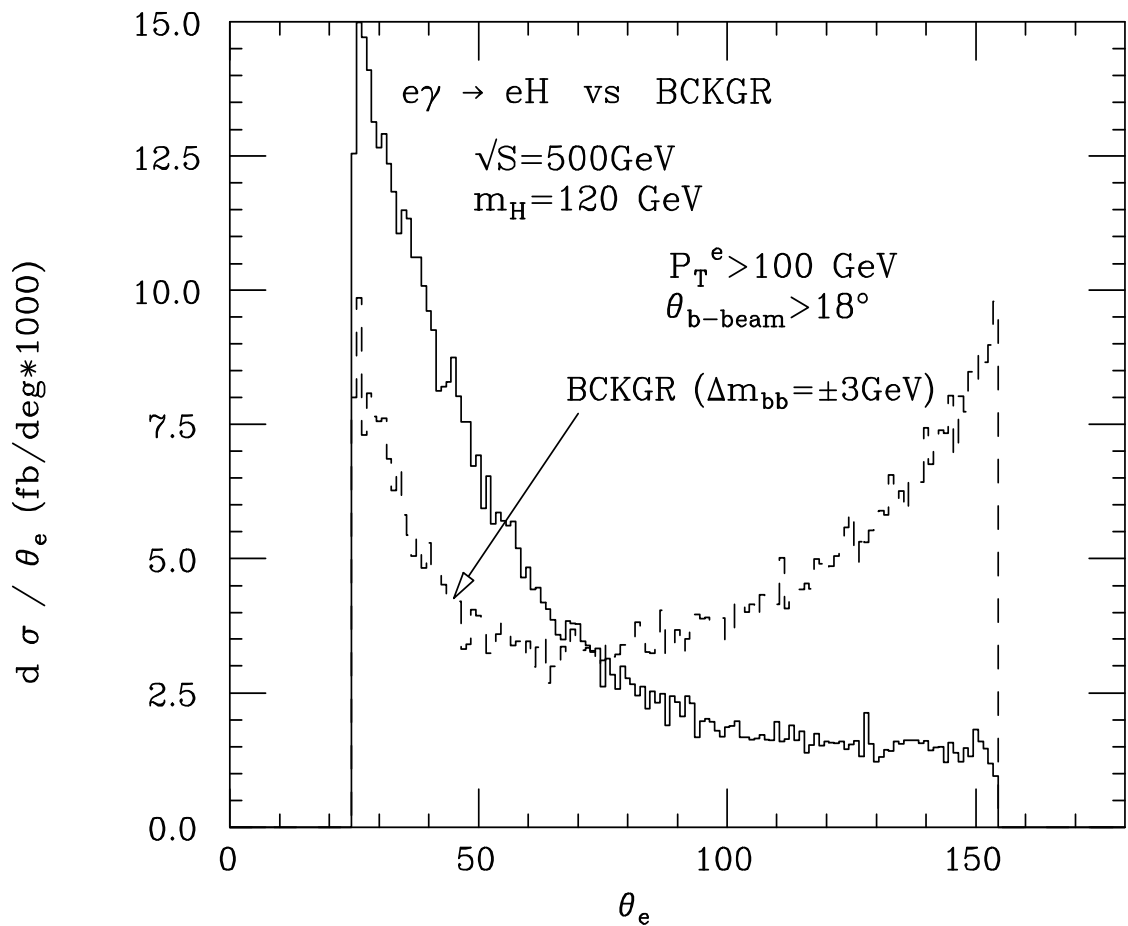,height=1.4\textwidth,
	width=\textwidth,angle=0}
}
\vspace*{-6.cm}
\caption{Final electron angular distribution with respect 
to the initial electron beam.
The solid (dashed) line refers to  the signal
(irreducible \eaebb background).
The kinematical cuts applied are shown in the plot.
The initial beams are assumed to be unpolarized.}
\label{fig71}
\end{figure}

Important improvements in the $S/B$ ratio can be obtained by
exploiting the final-electron angular asymmetry of the signal
(figure \ref{fig71}), by measuring the difference between the 
forward and backward cross sections.
A detailed analysis of this optimization procedure 
can be found in \cite{noii}.
With a luminosity of 100 fb$^{-1}$, at $\sqs=500$GeV,
one expects an accuracy as good as
about 10\% on the measurement of the \zah effects.
A luminosity of  50 fb$^{-1}$ would anyhow  allow to
measure the standard model signal with an accuracy better than 20\%.


\begin{table}[b]
\begin{center}
\begin{tabular}{|c||c|c|c||c|c|c|}
\hline
$m_H=120$~GeV
&\multicolumn{3}{c||}{$p^e_T > 100\,$GeV}
&\multicolumn{3}{c|}{$p^e_T > 10\,$GeV}\\
\cline{2-7}
$\sqrt s = 500$~GeV 
& $\sigma(e H)\,(fb)$ & $\sigma(e b\bar b)\,(fb)$ & $\sigma(e c\bar c)\,(fb)$ 
& $\sigma(e H)\,(fb)$ & $\sigma(e b\bar b)\,(fb)$ & $\sigma(e c\bar c)\,(fb)$ 
\\
\hline
$P_e = 0$  & 0.530  & 0.634 & 0.208 & 1.17  & 1.34 & 0.868  \\
\hline
$P_e = -1$ & 1.03   & 0.961 & 0.277 & 1.89  & 1.94 & 1.00  \\
\hline
$P_e = +1$ & 0.0249 & 0.304 & 0.136 & 0.422 & 0.767 & 0.726 \\
\hline
\end{tabular}
\end{center}
\caption{{\em  Comparison of the signal with the irreducible
background $e \gamma \to e b\bar b$ and the reducible background
coming from $e \gamma \to e c\bar c$, for different $e$-beam polarizations. 
For the $e \gamma \to e c\bar c$ background, a 10\% probability of 
misidentifying a $c$ quark into a $b$ is assumed
(that is, only 1/10 of the cross section is reported).
Two configurations for kinematical cuts are considered.
The angular cut $\theta(b(c)-beam)>18^o$ is  applied everywhere.
The signal rates includes 
the complete treatment of the $H\to b \bar b$ decay.
The $b\bar b$ invariant mass
for the background is integrated over the range 
$\mh-\delbb<\mbb(\mcc)<\mh+\delbb$, with $\delbb=3$GeV. 
}}
\label{tab41}
\end{table}

\begin{table}[t]
\vspace*{1cm}
\begin{center}
\hspace*{-1.5cm}
\begin{tabular}{|c|c||c|c|c|c|c|c|c|}
\hline
$\sqrt s =$& $m_H$
&\multicolumn{7}{c|}
{$\sigma(e\gamma\to e H, \;\; p^e_T > 100\,$GeV)$\;\;\;$(fb)} \\
\cline{3-9}
$500$GeV & (GeV) 
& Total & $|\gamma\gamma H|^2$& $|Z\gamma H|^2$ & $|$box$|^2$
& Int.$_{(\gamma\gamma H-Z\gamma H)}$  & Int.$_{(\gamma\gamma H-box)}$  
& Int.$_{(Z\gamma H-box)}$  
\\
\hline
$(P_e = 0 ;$ &
  80    &    0.618   &  0.264  &   0.146  &   0.0451 &
 0.0389& 0.0651& 0.0585 \\
$ P_{\gamma} = 0)$ &
  100    &    0.652  &   0.277  &   0.154  &   0.0481 &
0.0409 & 0.0695 &  0.0625 \\
&  120    &    0.705  &   0.296  &   0.166  &   0.0532 &
0.0439 & 0.0766 & 0.0692  \\  
&  140   &     0.818  &   0.341  &   0.190  &   0.0633  &
0.0505   & 0.0909  & 0.0822  \\
\hline
$(P_e = -1 ;$ &
  80    &    1.20   &  0.264  &   0.176   &  0.0899  &
0.423  &   0.124  &   0.122 \\
$ P_{\gamma} = 0)$ &
  100   &     1.27   &  0.277  &   0.185  &   0.0959  &
0.445  &   0.133  &   0.130 \\
&  120   &     1.37  &   0.296  &   0.199  &   0.106  &  
 0.478   &   0.147  &   0.144 \\
&  140   &     1.59  &   0.341  &   0.228   &  0.126  &  
 0.549  &  0.174  &   0.171 \\
\hline
$(P_e = +1 ;$ &
  80    &   0.0371  &0.264   &  0.117   & 3.55E-04&
 -0.345  &   5.86E-03& -4.94E-03 \\
$ P_{\gamma} = 0)$ &
  100    &    0.0386 & 0.277  &   0.123  &   3.66E-04&
 -0.363  &   6.21E-03& -5.22E-03 \\
&  120   &     0.0405 & 0.296  &   0.133   &  3.82E-04&
 -0.390  &   6.72E-03& -5.65E-03 \\
&  140    &    0.0465  &0.341  &   0.152   &  4.10E-04&
 -0.448 &    7.62E-03& -6.36E-03 \\
\hline
$(P_e = 0 ;$ &
  80     &   0.796   &  0.264   &  0.153   &  0.0798 &
0.118  &   0.0903  &0.0901  \\
$ P_{\gamma} = -1)$ &
  100    &    0.852  &   0.277   &  0.162   &  0.0860 &
 0.131  &   0.0986  &0.0983 \\
&  120   &     0.940  &   0.296  &   0.175  &   0.0962 &
 0.149  &   0.112 &    0.112\\
&  140   &     1.12  &   0.341  &   0.201  &   0.116  &  
 0.182  &   0.138  &   0.137 \\
\hline
$(P_e = 0 ;$ &
  80    &    0.440   &  0.264   &  0.139  &   0.0104 &
 -0.0403 & 0.0399 & 0.0268  \\
$ P_{\gamma} = +1)$ &
  100    &    0.452   &  0.277   &  0.146   &  0.0103 &
 -0.0487  &0.0403 & 0.0267 \\
&  120   &     0.470  &   0.296   &  0.157  &   0.0101 &
 -0.0610 & 0.0411 & 0.0268  \\
&  140     &   0.520   & 0.341  &  0.179  &   0.0101 &
 -0.0812 & 0.0434 & 0.0275  \\
\hline
\end{tabular}

\end{center}
\caption{{\em Interference pattern between the $\gamma\gamma Z$,
$Z\gamma H$ and boxes contributions versus the 
$e$-beam and $\gamma$-beam polarizations, for 
$p^e_T > 100\,$GeV. 
}}
\label{tab53b}
\vspace*{-.3cm}
\end{table}

\end{document}